\documentclass{iopart}
\usepackage{iopams}
\usepackage{graphicx}
\usepackage{color}

\begin{document}
\title[Optical layout for a 10\,m FP Michelson interferometer with tunable stability]{Optical layout for a 10\,m Fabry-P\'erot Michelson interferometer with tunable stability}

\author{Christian Gr\"{a}f\hspace{.3mm}$^{1}$, Stefan Hild$^{2}$, Harald L\"uck$^{1}$, \\Benno Willke$^{1}$, Kenneth A. Strain$^{2}$, Stefan Go\ss ler$^{1}$, \\and Karsten Danzmann$^{1}$}
\ead{christian.graef@aei.mpg.de}
\vskip 1mm
\address{$^{1}$\,Max--Planck--Institut f\"{u}r Gravitationsphysik (Albert-Einstein-Institut) and Leibniz Universit\"{a}t Hannover, Callinstr.~38, D-30167 Hannover, Germany}
\address{$^{2}$\,SUPA, School of Physics and Astronomy, The University of Glasgow, Glasgow, G12\,8QQ, UK}

\begin{abstract}
The sensitivity of high-precision interferometric measurements can be limited by Brownian noise 
within dielectric mirror coatings.  This occurs, for instance, in the optical resonators of gravitational 
wave detectors where the noise can be reduced by increasing the laser beam size.
 However, the stability of the resonator and its optical performance often impose a limit
on the maximally feasible beam size.  In this article we
describe the optical design of a 10\,m Fabry-P\'erot Michelson interferometer 
with tunable stability. Our design will allow us to carry out initial commissioning
with arm cavities of high stability, while afterwards the arm
cavity length can be increased stepwise towards the final,
marginally stable configuration. Requiring only minimal hardware changes, with respect to a comparable ``static'' layout,
the proposed technique will not only enable us to explore the stability limits of an optical 
resonator with realistic mirrors exhibiting inevitable surface imperfections, 
but also the opportunity to measure coating Brownian noise at frequencies as low 
as a few hundred Hertz. A detailed optical design of the
tunable interferometer is presented and requirements
for the optical elements are
derived from robustness evaluations. 
\end{abstract}


\section{Introduction}\label{sec:intro}
\begin{figure}[Htb]
\centering
\includegraphics[width=1\textwidth, clip=true, trim=0mm 0mm 10mm 0mm]{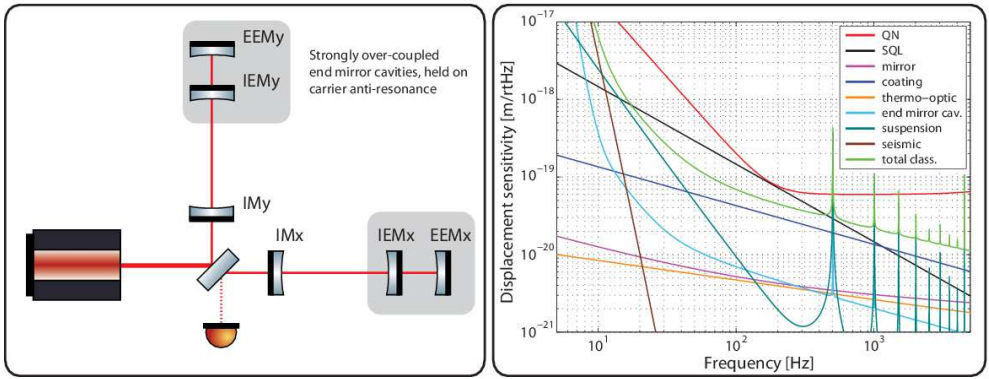}
\caption{\textbf{Left:} Schematic drawing of the AEI 10\,m sub-SQL interferometer target configuration, as presented in \cite{Kentaro10mDesign}. Topology-wise the interferometer is planned to be a Fabry-P\'erot Michelson Interferometer with an arm length of approximately
10\,m. The interferometer will not employ any recycling techniques. For the reduction of coating Brownian thermal noise it is intended to replace the conventional highly reflective arm cavity end mirrors with
short, strongly overcoupled cavities which will be held on anti-resonance for the carrier light by means of feedback control. \textbf{Right:} Anticipated noise spectral densities of components of the expected measurement noise for the target configuration. The total classical noise curve is 
the quadrature sum of residual seismic noise together with several thermal noise components: internal thermal noise (mirror), coating noise, thermo-optic noise and suspension thermal noise. An additional contribution to the total classical noise, specific to this design, stems from end mirror cavity-induced phase noise, cf. \cite{KhaliliCavity}.
In the most sensitive 
frequency band the total classical noise is below the sum of quantum radiation pressure and shot noise, thereby enabling a purely quantum noise-limited measurement of the differential interferometer arm length.  
In the frequency range where the quantum noise (QN) is close to the SQL, the dominant classical noise source is coating Brownian thermal noise which has to be at or below the design value in order not to mask quantum noise.}
\label{fig:SchematicSensitivityCombined}
\end{figure}

The AEI 10\,m Prototype \cite{AEI10mPT2010} is an ultra-low displacement noise facility, incorporating a large ultra-high vacuum system, excellent seismic isolation and a well-stabilized high-power laser source, and is intended to host a variety of interferometry experiments.
 One of these experiments is planned to be a Fabry-P\'erot Michelson interferometer which is intended to operate at a purely quantum noise limited sensitivity in its
detection band at hundreds of Hertz \cite{Kentaro10mDesign}. At a frequency of approximately 200\,Hz this instrument will be capable of reaching the standard quantum limit (SQL) of optical interferometry for 100\,g mirrors.
By creating quantum correlations within this interferometer, e.g.~by injecting squeezed vacuum, this limit can then even be surpassed. This will allow operating the interferometer at sub-SQL sensitivity, a state of operation which has to date not 
been reached by any interferometry experiment. A schematic drawing of the original interferometer conceptual design configuration, which in the following we will refer to as the ``target configuration'', is shown in figure~\ref{fig:SchematicSensitivityCombined}, along with the anticipated noise spectral densities.

It is evident that in the design of an instrument to reach the SQL, quantum noise must dominate over the sum of classical contributions which must be minimized.
The employment of advanced technologies, such as monolithic all-silica suspensions and ultra-low loss optics, as well
as a rigorous optimization of all relevant parameters is obligatory to reduce the individual types of thermal noise to a tolerable level. 
As in the case of large-scale advanced gravitational wave (GW) detectors, coating Brownian thermal noise is identified to be the most prominent classical noise source in the noise budget of the AEI 10\,m sub-SQL interferometer. 

Techniques have been developed to further increase the sensitivity of future GW detectors. These techniques, which are the subject of ongoing research, include 
modification of the optics (e.g.~TiO$_2$-doping of tantala/silica coatings \cite{TiDoping}, or the use of waveguides instead of dielectric 
mirrors \cite{Waveguide}), changes in the optical technologies (e.g.~interferometry with higher 
order optical modes such as the LG$_{33}$ mode \cite{HOM_Mours06, HOM_Chelkowski09}), and cryogenic cooling of the optics.  

This article reports on a stepwise approach to reducing coating thermal noise by iteratively enlarging the beam spots on the interferometer's 
arm cavity optics towards the technically feasible maximum. This goes hand in hand with pushing the arm 
cavities towards their geometric stability boundary. 

Typically, the radii of the laser beams on the interferometer optics are chosen much smaller than optics' radii to avoid excessive diffraction loss, and ensure stability of the optical mode. However, the larger the mirror surface area which is illuminated, the smaller 
the resulting coating thermal noise contribution. 
This reflects in the coating thermal noise theoretical model given in \cite{Harry2002}.

The use of extremely large laser beam spots is a key feature in the target configuration of the AEI 10\,m sub-SQL interferometer to reduce coating thermal noise below quantum noise level. This instrument is planned to be operated 
with beam spots with an equal radius of $w\approx9.7$\,mm on all cavity mirrors, which have a radius of
$a=24.3$\,mm. In this sense we regard our proposed setup as an intermediate, simplified configuration to pave the way to eventually building and operating our target configuration described in \cite{Kentaro10mDesign}.

\section{Challenges of stably operating the target configuration}\label{sec:challenges}
The attempt to operate a Fabry-P\'erot Michelson interferometer with extremely large beam spots on the cavity mirrors inevitably comes at the expense of poor resonator stability. The notion
of stability of an optical resonator is closely connected to the existence of low-loss cavity eigenmodes. 
With the aid of the formalism introduced in \cite{KogelnikLi} we can quantify the stability of an optical resonator as a function of the mirrors' radii of curvature and their spatial separation. This measure is commonly 
referred to as the cavities' \emph{g-factor}.
 
\subsection{Issues with marginally stable arm cavities}
The problem that arises from marginally stable (i.e.~$g \simeq 1$) optical resonators is that even small-scale length perturbations or mirror curvature error can render the instrument unstable. In an unstable resonator 
the property of self-consistency (i.e.~periodic re-focussing of the internal beam travelling back and forth between the mirrors) of stable cavities is violated. A substantial fraction of the 
light is therefore lost for the interferometric measurement, preventing the internal light field from fully building up, which would be crucial to reaching the design sensitivity of the interferometer. 
Furthermore, contrasting the case of stable resonators, heterodyne length control signals have been found to change their characteristics in marginally stable cavities \cite{EffectsOfModeDegeneracy}. 
This has implications for lock acquisition, as the transient signals present during that process are altered by spurious offsets which  arise when the cavity leaves the stable regime (e.g.~if a control actuator to adjust the 
cavity length imposes the tiniest amount of rotation on a mirror).  Even during a long lock, a brief disturbance could 
lead to instability which would produce offsets in the error signals of the control loops.  When this happens the control loops will command inappropriate corrections which may throw 
the system out of lock.

\subsection{Typical stabilities of the target configuration}
Our laboratory environment, basically the vacuum system, imposes space constraints on the minimum and maximum arm length of our interferometer.
In this respect, for the target configuration shown in figure~\ref{fig:SchematicSensitivityCombined}, typical arm cavity lengths are of the order $L_{arm} \approx 10.4\,$m. 
Another boundary condition with an impact on cavity lengths and mirror radii of curvature is the requirement for beam spots with a designated radius of $w=9.7$\,mm, which stems from a trade-off between low coating thermal noise and diffraction loss.
Obeying these boundary conditions, calculations
 yield an arm cavity g-factor of typically $g=0.999$\footnote{It must be noted that all stability estimates are approximate in the sense that they are based on the assumption of perfectly spherical optics. For more meaningful predictions of the 
stability, realistically imperfect optics need to be taken into account. In the experiment we propose to obtain surface maps of the actual mirrors, and to simulate the effect on stability. This is 
beyond the scope of this paper.}.
For such a configuration, a cavity length or a radius of curvature (RoC) error of only a few mm would be sufficient to render the cavity unstable.

For the purpose of comparison, typical arm cavity stabilities of large scale interferometric GW detectors and the AEI 10\,m sub-SQL interferometer are summarized in table~\ref{table:TypicalStabilities}.

\begin{table}[h!]
\footnotesize\rm
\caption{\label{table:TypicalStabilities} Comparison of arm cavity lengths and radii of curvature of cavity mirrors and the resulting cavity g-factors for large scale second and third generation GW observatories and the planned AEI 10\,m sub-SQL 
interferometer. Whereas second generation observatories exhibit a generous safety margin in their cavity stabilities, which is planned to be considerably smaller in the third generation detectors, the AEI sub-SQL interferometer 
arm cavities will eventually be operated extremely close to the stability boundary.}
\begin{tabular*}{\textwidth}{@{}l*{15}{@{\extracolsep{0pt plus12pt}}l}}
\br

&&\centre{2}{Radius of curvature}&\\
\ns\ns
&&\crule{2}&\\
& Cavity length &  Input mirror & End mirror  & Cavity g-factor\\
\hline\hline
Advanced LIGO \cite{aLIGO_OpticalParms} & 3996\,m& 1934\,m& 2245\,m & 0.832\\
Advanced Virgo \cite{AdVirgo_OpticalParms} & 3000\,m &  1420\,m & 1683\,m & 0.871\\
ET-B \cite{ET-B} &10000\,m&5070\,m&5070\,m & 0.945\\

\hline
\textit{Sub-SQL IFO simplified design:}\\
 Initial configuration  & 10.8\,m & 5.7\,m & 5.7\,m & 0.8\\
 Marginally stable configuration  & 11.3952\,m & 5.7\,m & 5.7\,m & 0.998\\
\hline
\br
\end{tabular*}
\end{table}

\section{Motivation for a stepwise approach towards the final beam size}\label{sec:motiStep}
In this article we propose
 a \emph{stepwise approach towards the final beam size} in order to ease the commissioning of the AEI 10\,m sub-SQL interferometer.
This approach will allow us to initially learn how to operate the interferometer with
relatively small beam spots on the cavity optics and therefore more comfortable arm cavity stability.
 After having established stable operation
in the initial configuration and gathering the required experience, 
we can then approach the marginally stable configuration by iteratively 
enlarging the beam size on the main mirrors towards its design
value.

For this stepwise approach to be feasible, it is crucial to  
find a way of increasing the beam size that does not require 
any major hardware changes, such as for instance replacing main optics.
It would for example be impractical and too cost intensive 
to adjust the beam size in the arm cavities by swapping the
main mirrors with ones with a different RoC, especially as the mirrors feature
monolithic suspension systems. 

\begin{figure}[Htb]
\centering
\includegraphics[width=1\textwidth]{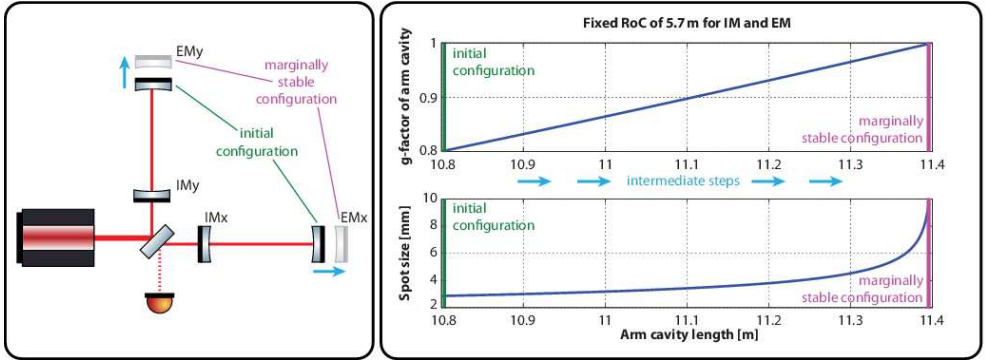}
\caption{Beam size and g-factor of a symmetric arm cavity of variable length 
with input mirror and end mirror curvature of the HR (high reflective-coated) surfaces of $R_{IM}^{HR}=R_{EM}^{HR}=5.7$\,m. The right end of the plot 
characterizes the marginally stable configuration of the AEI 10\,m sub-SQL interferometer which features extremely large 
beam spots and a g-factor close to instability. However, using exactly the same 
mirrors, but an arm cavity length shortened to $L_{arm}=10.8$\,m we can reduce the g-factor to
a comfortable value of $g=0.8$ while at the same time reducing the beam size of $w=9.72$\,mm to $w=2.86$\,mm.}
\label{fig:gfactor_vs_AClength}
\end{figure}

However, as the AEI Prototype infrastructure provides sufficient 
space to shift the positions of the main mirrors by up to about
1\,m or 10\,\% of the arm cavity length, we have the
possibility to reduce the beam size on the optics without adjusting
the main mirror RoC, but by initially
shortening the arm cavity length.
Let us assume design values for the arm cavity length of
$L_{arm}=11.395$\,m and radii of curvature of the input mirrors' (IM) and end mirrors' (EM) high reflective-coated (HR) surfaces of 
$R_{IM}^{HR}=R_{EM}^{HR}=5.7$\,m. Such an arm
cavity would have a g-factor of $g \approx 0.998$. As
shown in figure \ref{fig:gfactor_vs_AClength} we can achieve a comfortable g-factor of $g=0.8$
with exactly the same mirrors 
by just shortening the distance between the input and end
mirror by about 0.6\,m to a total arm cavity length of
$L_{arm}=10.8$\,m. Such a shortening of the arm cavity length
corresponds to reducing the beam size on the main 
mirrors from the targeted value of $w=9.72$\,mm to an initial beam
size of only $w=2.86$\,mm (see lower right subplot of figure \ref{fig:gfactor_vs_AClength}).  

\subsection{Direct measurement of coating Brownian noise}
\label{subsec:TN}

\begin{figure}[Htb]
\centering
\includegraphics[width=0.8\textwidth]{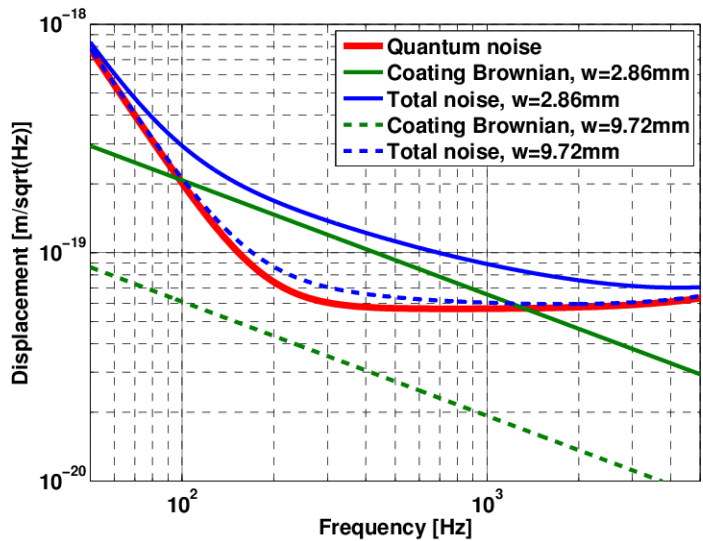}
\caption{Simplified displacement sensitivity graph for the AEI 10\,m sub-SQL interferometer only 
including the quantum noise and coating Brownian noise contributions for 
the initial configuration of 10.8\,m arm cavity length (solid traces) and 
the marginally stable configuration (dashed traces) with $L_{arm}=11.395$\,m arm cavity length.
The quantum noise (red) is independent of the arm cavity length.
With the initial configuration we expect to be able to directly
measure the coating Brownian noise in the frequency range between
100\,Hz and 1\,kHz, while in the marginally stable configuration with large 
beam spots thermal noise contributions will be
significantly below the quantum noise.
}
\label{fig:sensitivity_TN}
\end{figure}

Starting with the $L_{arm}=10.8$\,m configuration will not only be
advantageous for commissioning of the interferometer and
noise hunting, but will also allow us to directly measure coating
Brownian noise. Figure \ref{fig:sensitivity_TN} shows the
fundamental noise limits of the simplified AEI 10\,m sub-SQL interferometer design  
for the marginally stable configuration with $L_{arm}=11.395$\,m arm length and the initial configuration with 
$L_{arm}=10.8$\,m long arm cavities. Since the beam size on all main mirrors
is different by about a factor 3.5 between the two arm cavity lengths,
the coating Brownian noise will scale accordingly, thus
offering us  the possibility to directly 
measure coating Brownian noise at frequencies between about
100\,Hz and 1\,kHz with the initial configuration. This is an interesting opportunity to verify the coating Brownian noise
level at frequencies around 200\,Hz, which is the frequency range where
coating Brownian noise is most important for the advanced GW detectors,
and which has so far not been accessible by direct measurement
\cite{Numata03, Black04}.

\subsection{Potential impact on future gravitational wave detectors}
\label{subsec:impact2G}

One of the major steps for improving the sensitivity from 
the first to second generation GW detectors was
to significantly increase the beam size on the main test
masses, especially at the input mirrors. If larger mirror
substrates become available, future upgrades to these advanced
detectors might include even further increased beam
sizes on the mirrors in order to reduce the influence of
thermal noise contributions. This would require to operate 
the arm cavities with g-factors even higher than the ones 
stated in table \ref{table:TypicalStabilities}. 

The experience we will gain from the AEI 10\,m sub-SQL interferometer by step-wise approaching 
the cavity g-factor of $g=0.998$, will allow us to
study destabilizing effects and stability limitations
for related experiments. These results combined with reliable
simulations can be at least partially transfered to upgrades of second
generation GW detectors as well as to third generation 
GW detectors and may provide guidance in determining
  maximally feasible beam sizes
for these instruments.

\section{Properties of the optical configuration with tunable stability}\label{sec:properties}
Unlike the typical scenario in which Fabry-P\'erot Michelson interferometers are applied, 
in which the arm cavity geometry is not changed during the lifetime of the
experiment, the primary goal for the AEI 10\,m sub-SQL interferometer optical design is to identify a configuration which fulfills the requirement of tunable arm cavity stability or, 
synonymously, which can be operated equally well for different beam spot sizes on the cavity mirrors (cf. section \ref{sec:motiStep}). 

\begin{figure}[Htb]
\centering
\includegraphics[width=0.95\textwidth, clip=false, trim=0mm 0mm 10mm 0mm]{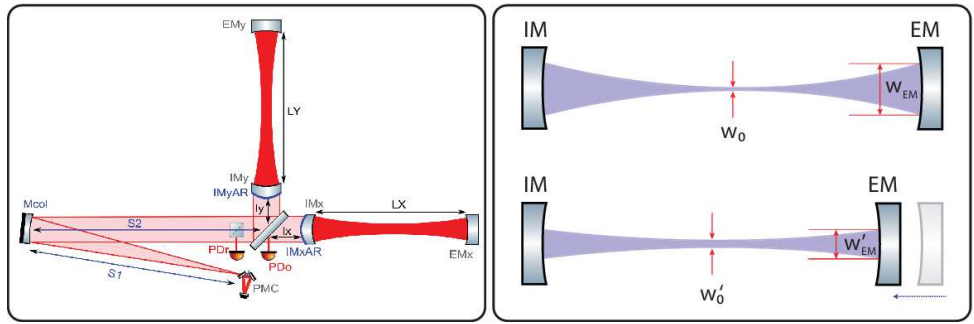}
\caption{\textbf{Left:} Simplified schematic drawing of the building blocks of the tunable cavity length configuration. The beam with a waist defined by the triangular pre-mode cleaner cavity (PMC) propagates to a
curved collimating mirror at a distance of about $12$\,m. The reflected beam is directed into the interferometer where it is matched to the arm cavities'
fundamental eigenmodes by means of curved arm cavity input mirror AR surfaces. \textbf{Right:} Moving the arm cavities' end mirrors alters the fundamental arm cavity eigenmode. For our starting configuration,
which features shorter arm cavities than the marginally stable design configuration, we find a larger beam waist at a shorter distance from the input mirror as well as smaller beam spots on both cavity mirrors.}
\label{fig:GaussSchematicEigenmodeCombined}
\end{figure}

Owing to the fact that each iteration step, at discrete
arm cavity lengths, will feature distinct cavity eigenmodes, it follows that the implementation of a flexible mode matching scheme is
the most elegant approach to solving this problem.
The performance goals of the instrument require close to optimal mode matching of the cavities at all times.

The starting point for our proposed optical configuration is the generation of a collimated laser beam with tunable radius.  This forms the 
input beam to the interferometer and is matched into the arm cavity eigenmodes by curved rear (antireflection coated) surfaces on the substrates 
of the input mirrors. 
A schematic drawing of this concept is depicted in the left pane of figure \ref{fig:GaussSchematicEigenmodeCombined}.

\subsection{Collimated interferometer input beam}
From the technical point of view, a collimated beam can easily be prepared by including a curved mirror into the input optics chain and by choosing the mirror's 
RoC and its distance to the input beam waist appropriately. 
To minimize the astigmatism introduced by the collimating mirror, the opening angle between the incident and the reflected beam should be as small as possible. This can be achieved by increasing the distance of beam propagation of the incoming and outgoing beam, e.g.~by 
positioning the collimating mirror near one of the interferometer arm cavity end mirrors.

Using a collimated input beam has a number of advantages over using a diverging beam\footnote{A possible downside may be a potentially higher susceptibility to beam pointing noise.
A detailed analysis of this effect is, however, not within the scope of this article.}. It is generally desired to have a high level of symmetry of the interferometer
arms because this has a high impact on the intrinsic cancellation of common mode perturbations at the beam splitter. On the other hand, to provide transmission of RF control
sidebands to the detection port of the interferometer, which is typically locked on or very close to a dark fringe, it is necessary to introduce a macroscopic offset 
in the path lengths between the two arm cavity input mirrors
and the beam splitter. This offset is referred to as the \emph{Schnupp asymmetry} \cite{Schnupp}. For a non-collimated input beam the propagation over unequal path lengths
would lead to beam parameters which were different on the parallel and the perpendicular arm cavity IMs. If perfect mode matching were to be
achieved for both arm cavities, this configuration would require either to include additional optics, or to have different radii of curvatures for the AR surfaces on the input mirrors.

A further benefit of using a collimated input beam is reduction of astigmatism introduced at the beam splitter.

\begin{figure}[Htb]
\centering
\includegraphics[width=1\textwidth]{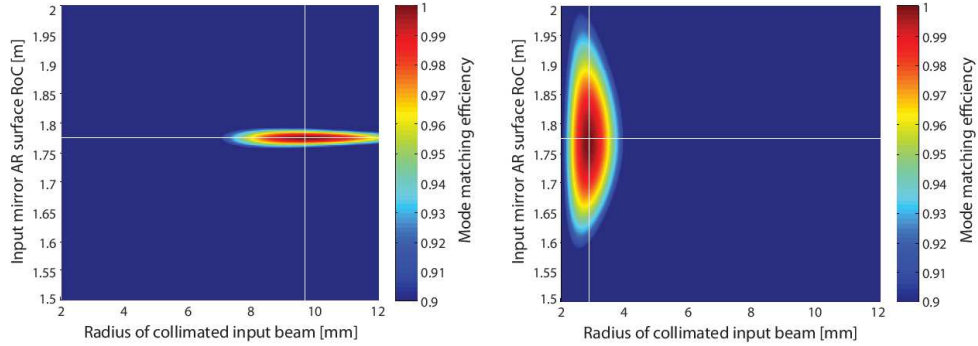}
\caption{Mode matching efficiency maps of the interferometer arm cavities. The left plot corresponds to the marginally stable configuration with
an arm cavity length of $L_{arm}=11.395$\,m, the plot on the right hand side illustrates the situation for the starting configuration with a reduced arm cavity 
length of $L_{arm}=10.8$\,m. In both maps the mode matching efficiency is color coded as a function of the radius of the collimated interferometer
input beam and the RoC of the AR surface of the arm cavity input mirrors. By holding the IM AR surface RoC constant and changing
the input beam radius only a mode matching efficiency for the short arm cavity configuration can be obtained which is degraded by approximately 1\,\% of
the (theoretically perfect) matching efficiency of the marginally stable configuration. All numerical investigations were carried out 
by means of the matrix formalism introduced in \cite{KogelnikLi} as well as the interferometer simulation software \emph{Finesse} \cite{FINESSE}.}
\label{fig:VisMapsCombined}
\end{figure}

\subsection{Coupling efficiency of the flexible mode matching scheme}
A natural measure to benchmark our proposed layout, especially with respect to the flexibility of the mode matching scheme, is the theoretical mode coupling efficiency for the 
extreme cases, i.e.~the initial configuration and the marginally stable 
configuration. 
The mode matching efficiency $\eta$, 
\begin{equation}
\eta = \frac{\left | \int \!\! \int\! dx\, dy \, \Psi(x,y,z) \, \hat{\Phi}^*(x,y,z) \right |^2}{\int\!\! \int\! dx\, dy \left | \Psi(x,y,z) \right |^2 \times \int\!\! \int\! dx\, dy \left | \hat{\Phi}(x,y,z) \right |^2} \quad ,
\end{equation}
which is referred to several times throughout this article, is a measure of the coupling of optical power into a fundamental cavity 
eigenmode. It is defined as the normalized overlap integral of the TEM$_{00}$ mode of the laser beam $\Psi$ and the fundamental cavity eigenmode $\hat{\Phi}$. 
The modes $\Psi$ and $\hat{\Phi}$ are fully determined by the complex beam parameters $q(z)$ and $\hat{q}(z)$ of the input beam and eigenmode at an arbitrary position $z=z_0$ 
along the beam axis of propagation.

The calculated efficiency can be regarded as an upper bound to the practically achievable mode matching quality.

We consider the marginally stable configuration as the reference, meaning that in our analysis all relevant parameters are chosen with respect to achieving perfect mode matching for this configuration. By keeping
all parameters, except for the arm cavity length, constant we can quantify the degradation of the mode matching and identify possibilities for the recovery of the mode matching as well as limits to the degree by which it is recoverable.

In our case, the mode matching efficiency is determined by two parameters, the radius of the collimated input beam as well as the RoC of the arm cavity input mirrors' AR surfaces. For the marginally stable configuration
with an arm cavity length of $L_{arm}=11.3952$\,m and mirror high reflectivity (HR) surface radii of curvature of $R_{IM}^{HR}=R_{EM}^{HR}=5.7$\,m we find an input beam radius of $w \approx 9.74$\,mm and an IM AR surface RoC of $R_{IM}^{AR} \approx 1.776$\,m to result in a perfect mode matching
to the arm cavities. 

Ideally, in the real interferometer the arm cavity length is changed in each iteration step by moving the end mirrors only; this we adopt as a further boundary condition for the stepwise cavity length tuning. If we now keep the optimal values of the marginally stable configuration for all parameters,  
except for the arm cavity length which we reduce to a value of $L_{arm}=10.8$\,m by shifting the end mirror towards the input mirror,
we observe a substantial decrease of the mode matching efficiency. This scenario corresponds to setting up the initial configuration with improved stability with optics that are optimized for marginally stable operation.

Owing to the fact that the RoC of the IM AR surfaces cannot be easily changed in practice, this value is to be considered a constant for all length iteration steps.
On the contrary, the radius of the collimated input beam can be tuned to recover the beam matching to the cavities' eigenmodes. According to this, by tuning the radius of the collimated input beam
down to $w \approx 2.88$\,mm the matching efficiency for the configuration with shortened arms can be $\eta > 99$\,\%, albeit the limitation of the radius of the collimated input beam being the only parameter available for optimization.
Aspects of the technical realization of a tunable collimated input beam are addressed in section \ref{sec:op_requirements}.

Based on experience gained from earlier experiments we consider a degradation of the mode matching efficiency of not more than 1\,\% (with respect
to the perfectly matched case) tolerable in the sense that this is likely to have a negligible impact on the performance of the instrument. A more elaborate estimation of this requirement based on
a detailed noise analysis is subject to future work.

The mode matching efficiency as a function of the radius of the collimated input beam and the RoC of the arm cavity input mirror AR surfaces for the two arm cavity length extremes is shown in figure~\ref{fig:VisMapsCombined}.

\begin{figure}[Htb]
\centering
\includegraphics[width=0.55\textwidth, clip=false, trim=0mm 0mm 10mm 0mm]{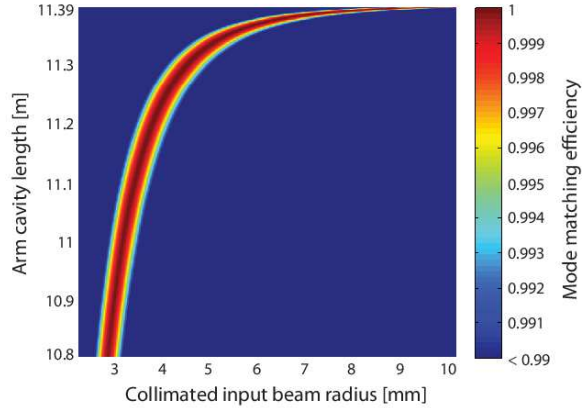}
\caption{Mode matching efficiency of the interferometer arm in the reflected port of the beam splitter (with respect to the input beam) as a function of the radius of the collimated input beam and the arm cavity length.
For the RoC of the arm cavity input mirror AR surface a value of $R_{IM}^{AR}=1.776$\,m was implicitly assumed. Whereas in theory perfect mode matching can be achieved for the marginally stable design 
configuration, a mode matching efficiency of up to $99$\,\% is theoretically feasible for the starting configuration with shorter arm cavities.}
\label{fig:MM_Map_DifferentLengths}
\end{figure}

The residual degradation in the short arm cavity case can be attributed to a waist position mismatch within the cavities, which cannot be compensated by tuning the input beam radius.
This is due to the fact that the focal distance of the curved IM AR surface for the collimated input beam is constant whereas the position of the waist of the arm cavity eigenmode is a function of the cavity length.
Due to the symmetry of the configuration, the waist position moves towards the IMs by half the length change. This is illustrated in the right pane in figure~\ref{fig:GaussSchematicEigenmodeCombined}.
The evolution of the mode matching efficiency as a function of the collimated input beam radius and the arm cavity length is shown in figure \ref{fig:MM_Map_DifferentLengths}. 

\subsection{Further operational requirements} \label{sec:op_requirements}
In our case, besides the introduction of the curved collimating mirror, the input optics chain needs to be extended by optics to implement the required feature of radius tunability of the collimated beam.
The notion of \emph{input optics} commonly summarizes the optical elements which serve the purpose to deliver a pure, well-aligned beam with the optimal geometry to the interferometer\footnote{For conciseness further optical elements such as electro-optic modulators, 
isolators, etc. which are usually required elements of the input optics chain, are omitted in this discussion.}.

It is possible to conceive various approaches to implementing adjustable modematching in the input chain.  
Obvious examples include exchanging the collimating mirror in each iteration step and introducing a beam expanding telescope in the collimated beam path, but these turn out to be poor choices. 
Whereas the former option depends on the time-consuming task of replacing a suspended optic and gives rise to a complicated re-alignment procedure in each iteration step, the
latter, likewise, requires frequent swapping of optics and may furthermore be an additional source of optical aberrations.  

Our preferred method of input beam shaping is to tune the waist radius of the \emph{initial} beam, while keeping the waist position constant, prior to its reflection at the collimating mirror. This can, for 
instance, be accomplished by means of a beam telescope in combination with a beam expander, which consist of lenses or mirrors. These can easily be 
shifted on the optical table for fine tuning. The use of active optics may help to avoid the need to exchange fixed focal length optical elements.

A matter closely related to the stable operation of the tunable length interferometer is the sub-area of sensing and control of the optical degrees of freedom of the instrument.
Typically, RF modulation based heterodyne length signal extraction schemes are employed, which require one or more electronic local oscillators as signal sources, whose frequencies are optimized with respect to 
cavity lengths within the interferometer to be controlled \cite{PDH}. In our case the cavity length tunability may require a flexible RF modulation scheme.

However, a detailed treatment of this topic, which can be regarded as a technical issue, rather than fundamental, is beyond the scope of this article and shall be discussed elsewhere.

\section{Estimation of the operational robustness}\label{sec:robustness}

Parameters in the optical layout may deviate from their designated values for a variety of reasons, e.g.~due to fabrication tolerances, environment-induced drifts 
or the nature of the experimental apparatus itself. 
Ideally, the interferometer design should exhibit a high level of immunity to tolerances in its constituting parameters. 
Practically we find imperfections in the optical elements and inaccuracies in the optical setup to degrade the performance of the interferometer or, in the worst case, to even render 
the instrument inoperable.

On the basis of the schematic shown in figure~\ref{fig:GaussSchematicEigenmodeCombined} we can identify design parameters which 
have a direct impact on the maximally achievable mode matching efficiency. These are: the initial beam
waist radius $w_0$ as well as its position $z_0$ (defined by the eigenmode of the triangular cavity in figure \ref{fig:GaussSchematicEigenmodeCombined}), the RoC of the 
collimating mirror as well as its position on the table and the RoC of the IMs' AR surfaces.

In this section we will investigate the impact of deviations of these parameters from their design values. 
This knowledge can in turn be utilized to formulate specifications for the required manufacturing precision for the optics.

\subsection{Initial laser beam waist radius and position}
Provision of an initial beam with well-defined beam parameters is crucial to meet the requirement for a well-collimated interferometer input beam with a specific radius for each cavity length iteration step. 
A mismatch of the actual parameters of the initial beam with respect to the ideal ones is likely to have a direct impact on the mode matching quality.

\begin{figure}[Htb]
\centering
\includegraphics[width=1\textwidth, clip=false, trim=0mm 0mm 0mm 0mm]{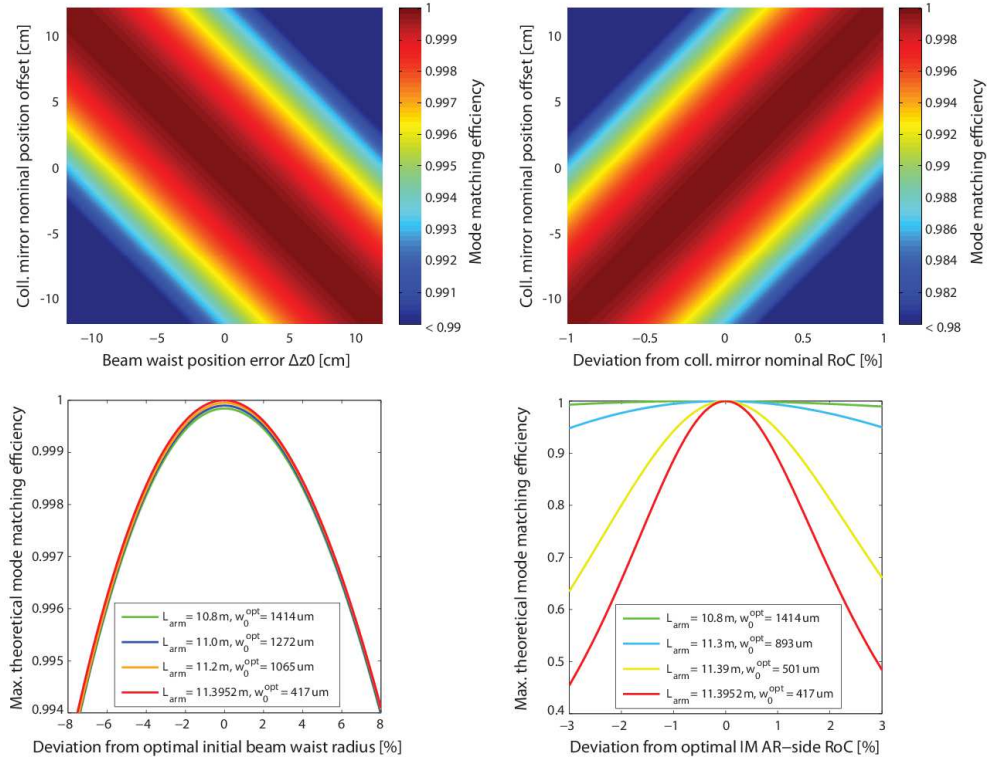}
\caption{Degradation of the theoretically achievable mode matching efficiency due to deviations of design parameters from their respective optima. \textbf{Top left:} A mismatch in the input beam waist position $z_0$ can be compensated 
by shifting the position of the collimating mirror on the table. \textbf{Top right:} Likewise, imperfections of the collimating mirror RoC can be 
compensated by shifting the mirror's position. \textbf{Bottom left:} 
The mode matching efficiency exhibits fairly low susceptibility to deviations from the optimal initial beam waist radius $w_0$. A deviation of $\pm 5$\,\% in $w_0$ results in a mode
matching efficiency degradation of less than 0.5\,\%. Note that none of the configurations, except for the marginally stable one, reaches perfect mode matching.
\textbf{Bottom right:} The susceptibility to IM AR surface RoC error increases with the arm cavity length approaching the marginally stable case. In the marginally stable configuration a deviation 
of $\pm 1\,\%$ comes at the expense of a mode matching efficiency degradation of $\approx 11.2\,\%$. }
\label{fig:2x2_Tolerancing}
\end{figure}

The dependence of the arm cavity mode matching efficiency on the initial beam waist position $z_0$ is depicted in the top left plot in
figure~\ref{fig:2x2_Tolerancing}. Clearly, a deviation of the waist position along the optical axis can be compensated by shifting the position of the collimating mirror by the same amount. 

It is noteworthy that shifting the collimating mirror position simultaneously alters the length of the incoming as well as the outgoing beam path.
Nevertheless, due to the reflected beam being collimated, this coupling of the two lengths is neutralized in first order. 
Consequently, the quality of the mode matching is mostly insensitive to length changes in this path. 

Small deviations of the initial beam waist radius from the optimum can be found to have a negligible effect on the mode coupling efficiency,
cf.~bottom left plot in figure~\ref{fig:2x2_Tolerancing}.
A deviation of $\pm 5$\,\% in $w_0$ results in a degradation of the mode matching efficiency of less than 0.5\,\%. 

\subsection{Collimating mirror RoC imperfections}
The effect of RoC imperfections of the collimating mirror as well as a possible workaround is illustrated in the top right plot in figure~\ref{fig:2x2_Tolerancing}. 

A RoC error results in a non-optimal focal length of the mirror. The focal length, in turn, is required to match the distance to the initial beam waist to perfectly collimate the beam in reflection.
Again, the collimating mirror position can be shifted to compensate this type of imperfection. The same argument of length offsets in the reflected beam path being negligible (see 
previous section) holds here, too. Alternatively, instead of shifting the mirror position, the RoC could e.g.~be thermally actuated upon. 

\subsection{Input mirror AR surface RoC}
The susceptibility of the mode matching efficiency to RoC imperfections of the IMs' AR surfaces is illustrated in the lower right plot in figure~\ref{fig:2x2_Tolerancing}. 
It becomes evident that whereas for the starting setup the arm cavity mode matching shows comparatively low susceptibility to this type of imperfection, the effect increases as we approach
the marginally stable configuration arm cavity length. While for the initial configuration it takes a RoC error of $\pm 3.2$\,\% to degrade the mode matching by $\approx 1$\,\%, for the marginally stable setup 
we find a RoC deviation of $\pm 1\,\%$ to result in a mode matching efficiency degradation of $\approx 11.2\,\%$. A mode matching efficiency of $\eta \gtrsim 99$\,\% for all configurations, including the marginally stable one, 
could be achieved by means of an IM AR surface RoC error lower than $\pm 0.28$\,\%, which corresponds to an absolute RoC error of $\pm 5$\,mm.

Unlike the cases discussed previously, for the IM AR surface RoC there is no well-decoupled degree of freedom in the instrument available that can be utilized to easily compensate an error in this parameter. 
Direct thermal actuation does not pose a suitable solution as the radii of curvature on both sides of the mirror would be affected simultaneously, leading to an unwanted distortion of the cavity eigenmode.
However, depending on its nature, a residual RoC error in both IM AR surfaces could be tackled by different means:

A ``common mode'' RoC error (i.e.~the sign of both RoC deviations, for the parallel and the perpendicular interferometer arm IM, is identical) of both IMs could be compensated by slightly tuning the divergence angle of the interferometer input beam. 
This could be achieved by means of shifting the collimating mirror from its optimal position or actuating on its RoC (e.g.~thermally). The pivotal point of this approach is to trade waist radius error 
for waist position error, the latter of which the cavity mode matching efficiency is generally less susceptible to. If, for instance, in the marginally stable configuration the actual IM AR
surface RoC turns out to be smaller by $1$\,\% with respect to its optimal value of $R_{IM}^{AR}=1.776$\,m, the mode matching efficiency can be recovered to $\eta \gtrsim 99$\,\% by increasing the RoC of the collimating mirror.
For typical beam path lengths in the collimating stage the required change of the collimating mirror RoC would be of the order of tens of centimeters.
Alternatively, the same can be achieved by shifting the initial beam waist out of the focal point of the collimating mirror, with an offset of the same order as the previously described collimating mirror RoC change.

A ``differential'' RoC error is in general harder to handle but 
could, if absolutely necessary, be compensated by introducing additional optical elements in the central Michelson arms, i.e.~between the beam splitter and the arm cavity IMs.

As mentioned previously, first and foremost these compensation techniques are relevant for configurations very close or at the marginally stable arm cavity length, only if
mirror AR surface RoC fabrication errors turn out to be larger than desired. For the larger part of the 
operation modes, in terms of differen cavity lengths, no such measures need to be taken.

\section{Summary and Outlook}\label{sec:summary}
In this article we have described a detailed optical layout
for the AEI 10\,m sub-SQL interferometer based on a robust procedure to bring the
interferometer to its final configuration with
marginally stable arm cavities. Starting with the arm cavities 
set to be shorter than eventually required, but with all other 
parameters unchanged, significantly increased stability of the 
arm cavity eigenmode may be obtained.
This is desirable to allow initial commissioning of the AEI 10\,m sub-SQL interferometer.  
A step-by-step approach to the final cavity mode is proposed.

In order to realize a close-to-optimal mode matching to the arm cavities, 
over the whole range of spot sizes, we employ
a collimated beam of variable size in combination with
input mirror substrates with curved front and rear sides. We found that the mode matching 
for different arm cavity lengths can be nearly completely 
recovered by changing the size of the incident laser beam,
while the associated change of the eigenmode waist position
only degrades the mode matching on the sub-percentage
level. 

The robustness analysis that was performed shows that 
the most stringent requirements for manufacturing accuracy
are imposed by the curvatures of the input mirror rear
surfaces, while deviations from all other design parameters are either mostly 
uncritical or can easily be compensated for by changing of free parameters. 

We have also pointed out that several aspects of the work presented 
in this article are of interest for the wider community,
such as for instance the possibility to directly measure
coating Brownian noise with the AEI 10\,m sub-SQL interferometer at frequencies around 200\,Hz. 
Moreover, the proposed optical layout will allow us to determine
how close to the instability one can realistically operate the arm cavities
of a Fabry-P\'erot Michelson interferometer, which is one of the
key-questions for future GW detectors. 

Future work will include the derivation of mirror polishing 
(and coating) requirements for the marginally stable arm cavities using 
numerical simulations with mirror maps. In addition to this, we
plan to analyze beam jitter requirements as well as laser 
noise couplings.

\ack{This work has been supported by the International Max Planck Research School (IMPRS)
and the cluster of excellence QUEST (Centre for Quantum Engineering and Space-Time Research).
S.~Hild and K.~A.~Strain are supported by the Science and Technology Facilities Council (STFC).}

\end{document}